\documentclass{emulateapj}

\makeatletter
\DeclareRobustCommand*\textsubscript[1]{%
  \@textsubscript{\selectfont#1}}
\def\@textsubscript#1{%
  {\m@th\ensuremath{_{\mbox{\fontsize\sf@size\z@#1}}}}}
\makeatother

\def\msh{{MSH~11--6\textit{2}}}

\newcommand\chandra{\textit{Chandra}}
\newcommand\xmm{\textit{XMM}}
\newcommand\fermi{\textit{Fermi}}
\newcommand\asca{{\em ASCA}}
\newcommand\be{\begin{equation}}
\newcommand\ee{\end{equation}}

\slugcomment{Accepted for Publication in ApJ}

\journalinfo{The Astrophysical Journal, in press}

\shorttitle{Broadband Emission from {\msh}}
\shortauthors{Slane et al.}

\begin{document}

\title{A Broadband Study of the Emission from the Composite Supernova
Remnant \msh}

\author{
Patrick~Slane,\altaffilmark{1}
John~P.~Hughes,\altaffilmark{2}
Tea~Temim,\altaffilmark{3}
Romain Rousseau,\altaffilmark{4}
Daniel Castro,\altaffilmark{1}
Dillon~Foight,\altaffilmark{1} 
B.~M.~Gaensler,\altaffilmark{5} 
Stefan~Funk,\altaffilmark{6}
Marianne~Lemoine-Goumard,\altaffilmark{4}
Joseph~D.~Gelfand,\altaffilmark{7} 
David~A.~Moffett,\altaffilmark{8} 
Richard~G.~Dodson,\altaffilmark{9}
and Joseph~P.~Bernstein\altaffilmark{10,11}} 

\altaffiltext{1}{Harvard-Smithsonian Center for Astrophysics, Cambridge, MA 02138-1516; slane@cfa.harvard.edu.}
\altaffiltext{2}{Dept. of Physics and Astronomy, Rutgers Univ., Piscataway, NJ 08854-8019; jph@physics.rutgers.edu.}
\altaffiltext{3}{NASA Goddard Space Flight Center - Code 662, Greenbelt, MD 20771; tea.temim@nasa.gov}
\altaffiltext{4}{Universit\'e de Bordeaux, Centre d'\'Etudes Nucl\'eaires Bordeaux Gradignan, CNRS-IN2P3, UMR 5797, Gradignan 33175, France; rousseau@cenbg.in2p3.fr, lemoine@cenbg.in2p3.fr}
\altaffiltext{5}{Sydney Institute for Astronomy, School of Physics A29, The University of Sydney, NSW 2006, Australia; bryan.gaensler@sydney.edu.au}
\altaffiltext{6}{Kavli Institute for Particle Astrophysics and Cosmology, Stanford Linear Accelerator Center, Stanford, CA 94025, USA; funk@slac.stanford.edu}
\altaffiltext{7}{New York University Abu Dhabi, P.O. Box 129188, Abu Dhabi, United Arab Emirates; jg168@astro.physics.nyu.edu}
\altaffiltext{8}{Dept. of Physics, Furman Univ., Greenville, SC 29613; david.moffett@furman.edu.}
\altaffiltext{9}{International Centre for Radio Astronomy Research
University of Western Australia; richard.dodson@icrar.org.}
\altaffiltext{10}{High Energy Physics Division, Argonne National Laboratory, Argonne, IL 60439; jpbernst@anl.gov.}
\altaffiltext{11}{Department of Astronomy, University of Michigan, Ann Arbor,
MI 48109}

\begin{abstract}
\msh\ (G291.1$-$0.9) is a composite supernova remnant for which
radio and X-ray observations have identified the remnant shell as
well as its central pulsar wind nebula. The observations suggest a
relatively young system expanding into a low density region. Here
we present a study of \msh\ using observations with the \chandra,
\xmm, and \fermi\ observatories, along with radio observations from
the Australia Telescope Compact Array (ATCA). We identify a
compact X-ray source that appears to be the putative pulsar that
powers the nebula, and show that the X-ray spectrum of the nebula
bears the signature of synchrotron losses as particles diffuse into
the outer nebula.  Using data from the \fermi\ LAT, we identify
$\gamma$-ray emission originating from \msh.  With density constraints
from the new X-ray measurements of the remnant, we model
the evolution of the composite system in order to constrain the
properties of the underlying pulsar and the origin of the $\gamma$-ray
emission.
\end{abstract}

\keywords{ radiation mechanisms: non-thermal
--- ISM: supernova remnants: --- ISM : individual (MSH~11--6\textit{2})
--- stars: neutron}

\section{Introduction}

The creation of an energetic pulsar in a supernova explosion often
results in the formation of a composite supernova remnant (SNR)
that can be characterized by emission from the remnant, an associated
pulsar wind nebula (PWN), and the neutron star (NS) itself (see,
e.g., Gaensler \& Slane 2006).  Recent studies of such systems have
resulted in discoveries of new pulsars, constraints on the spectra
of particles injected from the pulsars into their nebulae, and
evidence for interactions between the SNR reverse shock (RS) and
the PWN.  Broadband observations of composite SNRs have been crucial
for improving our understanding of the detailed evolution of these
systems, particularly with the vast improvements in capabilities
provided by current X-ray and $\gamma$-ray observatories.

\msh\ (Mills, Slee, \& Hill 1961) is a centrally-brightened SNR whose
properties strongly suggest the presence of a central nebula powered
by an active pulsar.  Radio observations reveal a remnant with a
distinct bar-like central structure whose flat radio spectrum
($\alpha_r = 0.29 \pm 0.05$, where $S_\nu \propto \nu^{-\alpha_r}$)
is similar to that observed for PWNe, and polarization measurements
at 5 and 8.4~GHz indicate that the magnetic field in the central
region of the SNR is aligned with the long axis of the bar (Roger
et al. 1986); the flux density at 8.4~GHz is $10.4 \pm 0.4$~Jy.  No
radio pulsar associated with \msh\ has been detected.  An 11~hr
observation at 1.4~GHz failed to detect any pulsations from this
region, down to a sensitivity limit of 0.028 mJy assuming a 10\%
duty cycle\footnote{F. Camilo, private communication.}.

The distance to \msh\ is poorly known. Limits on HI absorption
indicate a distance greater than 3.5 kpc (Moffett, Gaensler, \&
Green 2001; Moffett et al. 2002).  We adopt a fiducial distance of
5~kpc in derived quantities below, and scale values to $d_5$, the
distance in units of 5~kpc.  
%Based on the radio synchrotron spectrum,
%the minimum-energy magnetic field in the central nebula is $B \sim
%65 d_5^{-2/7} \mu{\rm G}$ (Harrus, Hughes, \& Slane 1998). 
Based on the radio synchrotron spectrum,
the minimum-energy magnetic field in the central nebula has been
estimated as $B \sim
65 d_5^{-2/7} (\theta_P/2^\prime.5)^{-6/7}  \mu{\rm G}$, where
$\theta_P$ is the PWN radius in arcmin (Harrus, 
Hughes, \& Slane 1998). 

X-ray emission from \msh\ was first identified by Wilson (1986)
using the {\em Einstein Observatory.} X-ray observations with the
\asca\ observatory reveal a spectrum best described by a combination
of thermal and nonthermal models (Harrus, Hughes, \& Slane 1998).
The hard X-ray emission is spatially concentrated in the central
regions, providing further evidence for the presence of a PWN within
the remnant, but the limited angular resolution of \asca\ precluded
detection of the pulsar itself.  The derived spectral index for the
X-ray power law component is $\Gamma = 2.0_{-0.3}^{+0.1}$ and the
associated luminosity is $L_x (0.2-4 {\rm\ keV}) \sim 6 \times
10^{33} d_5^2 {\rm\ erg\ s}^{-1}$.  The best-fit temperature of the
thermal component, presumably associated with emission from
shock-heated gas in the SNR shell, is $kT \sim 0.8 {\rm\ keV}$.
These observations clearly imply that \msh\ harbors a young pulsar
that powers a central PWN. \chandra\ observations, which we discuss
in detail below, reveal a compact X-ray source that is almost
certainly the pulsar counterpart (Harrus et al. 2002; Kargaltsev
\& Pavlov 2008).

Observations with the {\em Compton Gamma-Ray Observatory} discovered
the source 3EG~J1102$-$6103, for which \msh\ falls within the
position error circle. The flux is $F(>100 {\rm\ MeV}) = (3.3 \pm
0.6) \times 10^{-7}{\rm\ photons\ cm}^{-2}{\rm\ s}^{-1}$, with a
power law index $\Gamma = 2.5 \pm 0.2$ (Hartman et al. 1999).
This source was suggested as a $\gamma$-ray counterpart to \msh\
(Sturner \& Dermer 1995), but was also subsequently suggested as
the counterpart to the nearby young pulsar J1105$-$6107 (Kaspi et
al. 1997). Given the large uncertainty in the position of the
$\gamma$-ray source, the origin of its emission remained uncertain.

Here we report on observations with the ATCA, the {\em Chandra X-ray
Observatory}, the {\em XMM-Newton X-ray Observatory}, and the {\em
Fermi Gamma-Ray Space Telescope} that provide sufficient sensitivity
and resolution to identify the neutron star in \msh, and to provide
a much-improved characterization of the properties of this composite
system. In Sections 2 we discuss our ATCA observations.
In Section 3 we summarize the \chandra\ and \xmm\ observations,
with detailed investigations of the PWN, neutron star, and SNR
shell.  In Section 4 we discuss observations with the {\em Fermi
LAT}, and in Section 5 we present discussions of the results along
with our conclusions.

\section{Radio Observations}

\subsection{Observations and Data Reduction}

%%%%%%%%%%%%%%%%%%%%%%%%%% Figure 1 %%%%%%%%%%%%%%%%%%%%%%%%%%%%%%
\begin{figure}[t]
\centerline{\includegraphics[width=3.4in]{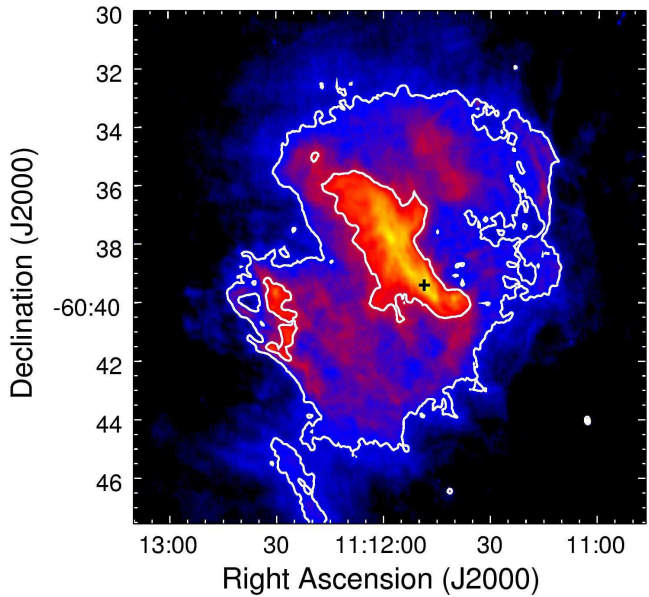}}
\caption{ATCA 1.4 GHz image  of \msh. The remnant is characterized by
a faint shell with a bright central bar. Contours levels are 3.6
and $1.5 {\rm\ mJy\ beam}^{-1}$. The cross marks the position of
the X-ray point source described in Section 3.}
\end{figure}
%%%%%%%%%%%%%%%%%%%%%%%%%%%%%%%%%%%%%%%%%%%%%%%%%%%%%%%%%%%%%%%%%%

Radio observations of \msh\ were carried out with the
ATCA, a 6-element interferometer located near Narrabri, New South
Wales, Australia.  Observations were carried out on three separate
dates using three different array configurations: on 1999 Feb 10
(750C array), 1999 Feb 22 (6C array), and 1999 Mar 24 (1.5B array).
In all cases, radio continuum data were recorded in all four
polarizations, over a bandwidth of 128~MHz centered at 1384~MHz.
At each epoch, \msh\ was observed for 8--11 hours spread
over a 12-hour synthesis, combined with regular short observations
of the extragalactic source PKS~B1036--697 to calibrate atmospheric
gain variations and polarization leakages. The bright source
PKS~B1934--638 was also observed at each epoch to provide an overall
flux calibration.

Following editing of the data and calibration in the {\tt MIRIAD}\
package using standard techniques, the visibility data-set was
imaged using multi-frequency synthesis and uniform weighting, and
then deconvolved with a maximum entropy algorithm. The final total
intensity image has an angular resolution of $8\farcs8 \times
6\farcs6$, with the major axis oriented at a position angle $3^\circ$
east of north, and an RMS sensitivity of 0.1--0.2~mJy~beam$^{-1}$.
Preliminary investigations of these observations have been presented
by Moffett, Gaensler, \& Green (2001) and Moffett et al. (2002).

\subsection{Radio Structure and Properties}

%%%%%%%%%%%%%%%%%%%%%%%%%% Figure 2 %%%%%%%%%%%%%%%%%%%%%%%%%%%%%%
\begin{figure}[t]
\centerline{\includegraphics[trim=0.8in 0.8in 0.8in
1.2in,clip,angle=270,width=3.9in]{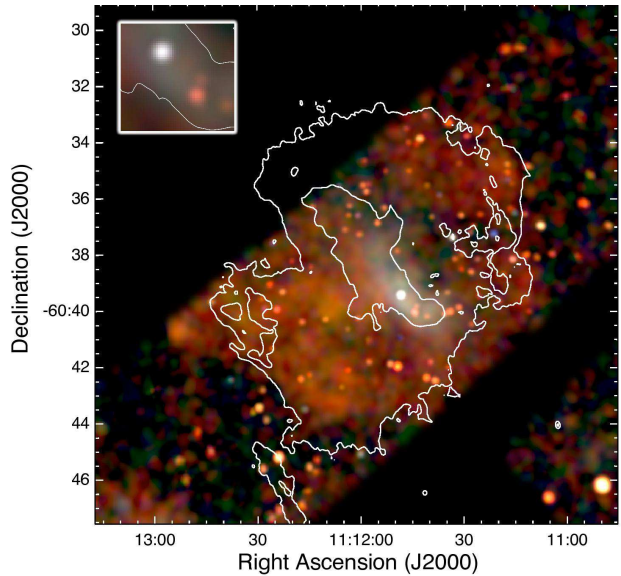}}
\caption{
\chandra\ image of \msh. Colors correspond to different X-ray energy
bands: red = 0.75 - 1.45 keV; green = 1.45 - 2.58 keV; blue = 2.58
- 6.0 keV.  The SNR shell is evident as soft (red) X-ray emission
(the portion of the shell in the northeast falls outside the detector
field of view). The central bar and associated compact source are
white in color, corresponding to harder emission.  The inset figure
shows the central regions of the PWN containing the pulsar candidate
(white) and the star Tycho 8959 255 1 (red).
}
\end{figure}
%%%%%%%%%%%%%%%%%%%%%%%%%%%%%%%%%%%%%%%%%%%%%%%%%%%%%%%%%%%%%%%%%%

The ATCA 1.4 GHz image of \msh\ (Figure 1) reveals a roughly circular
morphology with a bright central bar-like structure that corresponds
with the flat-spectrum region identified by Roger et al. (1986).
Partial limb-brightening is observed along the northwestern and
southeastern boundaries of the SNR shell.  Considerable filamentary
structures are observed both along the limb and in the SNR interior.
The bright central bar reveals several regions of enhanced radio
emission, with the brightest feature located near the southwestern
end of the structure (at an approximate position of 11$^{\rm
h}$11$^{\rm m}$48$^{\rm s}$,-60$^\circ$39$^\prime$27$^{\prime\prime}$).
The structure of the central bar is similar to that observed in
several other PWNe, including G11.2$-$0.3 (Tam et al.  2002), N157B
(Lazendic et al. 2000), and G328.4$+$0.2 (Johnston, McClure-Griffiths,
\& Koribalski 2004; Gelfand et al. 2007).  Its apparent similarity
to the PWN in G328.4$+$0.2, in particular, may suggest that the
nebula has been compressed by the SNR reverse shock (RS), which may
have propagated more quickly from the NW/SE directions, where the
SNR limbs show slight brightness enhancements.

The total 1.4~GHz flux from the SNR is 14 Jy, with 4.8 Jy
associated with the central nebula. 
This revised flux value results
in a new minimum-energy estimate for the magnetic field:
$B \approx 36 d_5^{-2/7} (\theta_P/2^\prime.5)^{-6/7}  \mu{\rm G}$.
As we discuss in Section 5.3, however, some evolved PWNe are known
to be particle-dominated, with magnetic field strengths far below the
minimum-energy (or equipartition) values.

\section{X-ray Observations}

A 50 ks \chandra\ ACIS observation of \msh\ was performed on 2002
April 8 (ObsID 2782). The central region of the SNR was placed on
the S3 chip of the detector, and portions of the shell were imaged
on the adjacent chip S2. The extent of the SNR (with a radius $R_{\rm
SNR}  \approx 5.7$~arcmin) is considerably larger than the field
covered by the S3 chip, which resulted in the northeastern regions
of the remnant falling outside of the detector field of view.

\msh\ was observed for 40 ks with \xmm, on 2002 February 6 (ObsID
0051550101). The MOS2 and pn detectors were operated in full frame
mode with the MEDIUM filters, providing full coverage of the SNR,
while MOS1 was operated in timing mode with the THIN filter.

\subsection{Observations and Data Reduction}

The \chandra\ data were reduced using standard screening prescriptions.
All data were reduced using tools from {\it CIAO 3.2.2.}. After
screening, the remaining exposure time was 49.4 ks.

The \chandra\ observations show hard emission extended along the
radio bar (Figure 2), although the emission does not extend as far
to the NE as the radio bar.  There is also a clear shell of soft X-ray
emission that corresponds well with the morphology of the radio
SNR, including a small intensity enhancement along the position of
the western radio limb-brightening, and a large underluminous region
to the northwest of the radio bar.

Of particular significance in the \chandra\ image is the presence of 
numerous point sources of X-ray emission along the line of sight
to \msh. These appear to be associated with the open cluster Tr~18,
located at a distance of $\sim 1.5$~kpc (Vazquez \& Feinstein 1990).

We have analyzed spectra from both the central X-ray bar and the
SNR shell using background regions selected from regions adjacent to
the bar, and just outside boundaries of the remnant, respectively.
Weighted response matrix (RMF) and ancillary response (ARF) files
were created for the extended regions, and spectral modeling was
then carried out using {\sc xspec} version 11.3.2ag. Point sources
within in the regions of interest were excluded from the spectral
extraction.

The \xmm\ data were reduced using SAS Version 10.0. Standard screening
resulted in cleaned exposures of 35.6 (29.2) ks for MOS2 (pn).

\subsection{Central Bar}

The central X-ray bar is aligned with the radio bar, and extends
to the edge of the radio in the southwest, but the brightness falls
off more quickly than the radio emission in the northeast direction.
There is a bright X-ray source located at the position of the
brightest radio emission, likely representing the pulsar that powers
the nebula. The source, located at 11$^{\rm h}$11$^{\rm m}$48$^{\rm
s}$.62,-60$^\circ$39$^\prime$26$^{\prime\prime}$.2 (and hereafter
identified as CXOU J111148.6-603926), is roughly centered on the
X-ray nebula. A fainter point source is detected at the southwestern
edge of the radio bar, with coordinates 11$^{\rm h}$11$^{\rm
m}$44$^{\rm s}$.5,-60$^\circ$40$^\prime$02$^{\prime\prime}$.8 (see
inset to Figure 2).  The source is positionally coincident with the
$V = 10.8$ star Tycho 8959 255 1, which is unrelated to the nebula.

The \chandra\ spectrum of the X-ray bar (excluding CXOU J111148.6$-$603926)
is well-described by a power law model with $N_H \approx 7 \times
10^{21}{\rm\ cm}^{-2}$ and $\Gamma \approx 1.7$ (Table 1, Figure
3), securing the PWN interpretation for this structure.  Here we
have used an elliptical background region on S3, adjacent to the
PWN but within the SNR, to account for both sky and instrumental
background as well as thermal emission from the SNR.  An independent
fit using the \xmm\ pn data yields virtually identical results. The
luminosity ($0.5 - 10$~keV) is $1.1 \times 10^{34} d_5^2 {\rm\ erg\
s}^{-1}$, about a factor of two higher than the value determined
from \asca\ observations (Harrus, Hughes, \& Slane 1998), but far
lower than the value of $\sim 8 \times 10^{34} d_5^2 {\rm\ erg\
s}^{-1}$ cited by Kargaltsev \& Pavlov (2008).

%%%%%%%%%%%%%%%%%%%%%%%%%% Figure 3 %%%%%%%%%%%%%%%%%%%%%%%%%%%%%%
\begin{figure}[t]
\centerline{\includegraphics[width=3.4in]{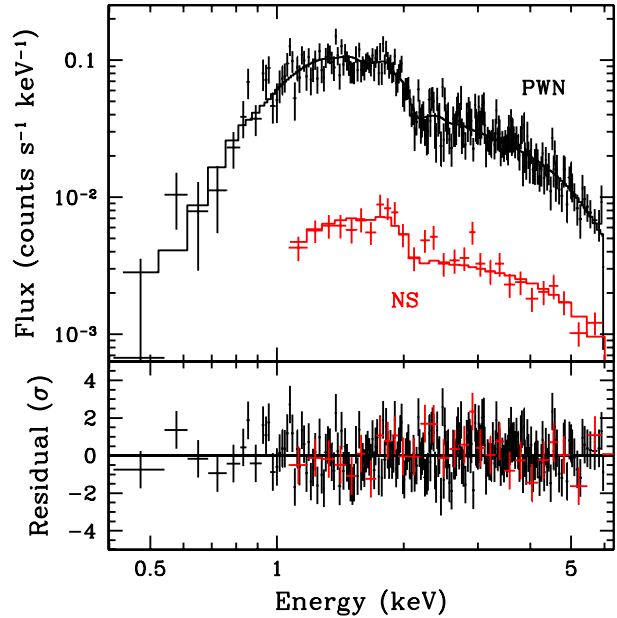}}
\caption{
{\chandra} ACIS X-ray spectra for PWN (black) and NS (red)
with absorbed power law fits.
}
\end{figure}
%%%%%%%%%%%%%%%%%%%%%%%%%%%%%%%%%%%%%%%%%%%%%%%%%%%%%%%%%%%%%%%%%%

%%%%%%%%%%%%%%%%%%%%%% Table 1 %%%%%%%%%%%%%%%%%%%%%%%%%%%%%%%%

\begin{deluxetable}{ccccc}
\tablecolumns{4}
\tabletypesize{\scriptsize}
\tablewidth{0pc}
\tablecaption{Spectral Fit Parameters}
\tablehead{
\colhead{Region} &
\colhead{Data} &
\colhead{Parameter} &
\colhead{Value}
}
\startdata
PWN & CXO & $N_H$ & $(6.7 \pm 0.7) \times 10^{21}{\rm\ cm}^{-2}$ \\
(whole) && $\Gamma$ & $1.8 \pm 0.1$ \\
&& $F_x^a$ & $3.5 \times 10^{-12}{\rm\ erg\ cm}^{-2}{\rm\ s}^{-1}$ \\
&&& \\
PWN & CXO & $N_H^a$ & $6.7 \times 10^{21}{\rm\ cm}^{-2}$ \\
Region 1 && $\Gamma$ & $1.7 \pm 0.1$ \\
Region 2 &&& $ 1.8 \pm 0.1$ \\
Region 3 &&& $2.2 \pm 0.1$\\
Region 4 &&& $1.5 \pm 0.1$\\
Region 5 &&& $1.7 \pm 0.1$ \\
Region 6 &&& $2.5 \pm 0.4$\\
&&& \\
NS & CXO & $N_H^a$ &  $6.7 \times 10^{21}{\rm\ cm}^{-2}$ \\
&& $\Gamma$ & $1.2 \pm 0.2 $\\
& & $F_x^b$  & $3.7 \times 10^{-13}{\rm\ erg\ cm}^{-2}{\rm\ s}^{-1}$ \\
&&& \\
SNR  & XMM & $N_H^a$ & $6.7 \times 10^{21}{\rm\ cm}^{-2}$ \\
(whole) & & $kT_1$ & $1.3^{+0.5}_{-0.2}$~keV\\
& & $n_e t$ & $(1.5 \pm 0.2) \times 10^{10} {\rm\ s\ cm}^{-3}$\\
& & $kT_2$ & $2.8 \pm 0.4$~keV\\
& & $F_1^b$ & $2.1 \times 10^{-11}{\rm\ erg\ cm}^{-2}{\rm\ s}^{-1}$ \\ \\
& & $F_2^b$ & $3.0 \times 10^{-12}{\rm\ erg\ cm}^{-2}{\rm\ s}^{-1}$ \\ \\
\enddata
\tablecomments{\\
a) Fixed at best-fit value for PWN\\
b) Unabsorbed 0.5 - 10.0 keV flux. \\
}
\end{deluxetable}

%%%%%%%%%%%%%%%%%%%%%%%%%%%%%%%%%%%%%%%%%%%%%%%%%%%%%%%%%%%%%%%%%%%%%%%%

To search for spectral variations in the PWN, we extracted spectra
from rectangular regions along the nebula, as shown in Figure 4.
We fixed the column density to that obtained for the entire PWN,
and then fit each spectrum to a power law model. The results,
summarized in Table 1, show a spectral steepening with distance
from the central neutron star. Such steepening is also observed in
G21.5$-$0.9 (Slane et al. 2000), 3C~58 
(Bocchino et al. 2001; Slane
et al. 2004) and other PWNe, and is understood to be associated
with synchrotron aging of the particles as they diffuse from the
injection point to the outer nebula (Kennel \& Coroniti 1984),
although models matching the exact radial behavior of the spectral
index remain elusive (e.g. Reynolds 2003).

%%%%%%%%%%%%%%%%%%%%%%%%%% Figure 4 %%%%%%%%%%%%%%%%%%%%%%%%%%%%%%
\begin{figure}[t]
\centerline{\includegraphics[width=3.4in]{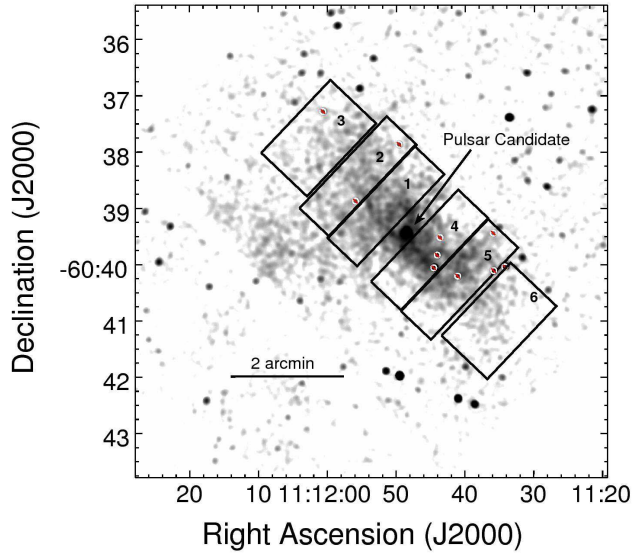}}
\caption{
\chandra\ image of PWN in \msh, along with regions used for extracting
the spectral index along the nebula (see Table 1). The image has been
smoothed with a Gaussian of 5~arcsec width. The pulsar candidate
is the bright central source between regions 1 and 4. White circles
with red slashes indicate point sources removed before extraction
of spectra.
}
\end{figure}
%%%%%%%%%%%%%%%%%%%%%%%%%%%%%%%%%%%%%%%%%%%%%%%%%%%%%%%%%%%%%%%%%%

\subsection{Compact Source}

Our \chandra\ observations obtain 820 background-subtracted counts
from within a circle with a radius of 3.3 arcsec centered on the
compact source CXOU J111148.6-603926.  The associated count rate
of $1.6 \times 10^{-2} {\rm\ cnt\ s}^{-1}$ yields negligible pileup
in the image. The radial profile of the compact source image is
well-described by the point spread function of the telescope, with
no evidence of a surrounding jet/torus structure as is seen in many
other such systems, although evidence for filamentary structure
is seen on larger scales (see Figure 4).

The spectrum of the compact source is well-described by a power law
with the column density fixed at the value obtained from fits of
the entire PWN, and with $\Gamma = 1.2
\pm 0.2.$ The unabsorbed flux is $F_x(0.5-10 {\rm\ keV}) = 3.7 \times
10^{-13}{\rm\ erg\ cm}^{-2}{\rm\ s}^{-1}$ indicating a luminosity
$L_x (0.5-10 {\rm keV}) = 1.1 \times 10^{33} d_5^2 {\rm\ erg\ s}^{-1}$,
similar to that measured for a number of other young pulsars with
bright PWNe (see Kargaltsev \& Pavlov 2008).

\subsection{SNR Shell}

To investigate the X-ray properties of the SNR shell, we extracted
spectra of the entire SNR from both the \xmm\ pn and MOS2 detectors
(see Figure 5).  Emission from the PWN was eliminated from the
spectra, as was emission from point sources in the field.
Weighted response files were generated for each spectrum, and
background spectra were obtained from regions outside the SNR shell
in each detector.

The spectra (Figure 6) were fit simultaneously to an absorbed
nonequilibrium ionization planar shock model ({\sc vpshock} in {\sc
xspec}), with the column density fixed at the best-fit value for
the PWN. We find an electron temperature $kT = 1.5^{+0.5}_{-0.2}$~keV
and an ionization timescale $n_e t = (2.9 \pm 0.9) \times 10^{10}
{\rm s\ cm}^{-3}$, with evidence for mild overabundances of Ne, Mg,
and Si. The fit requires a second, harder component, which can be
fit with a power law with $\Gamma = 2.6 \pm 0.2$ ($\chi^2_r = 1.3$
with 612 degrees of freedom), but a slightly better fit ($\Delta
\chi^2 = 6.1$ for the same number of degrees of freedom) is obtained
with a second thermal component ({\sc nei} model in {\sc xspec})
with $kT = 2.8 \pm 0.4$~keV and cosmic abundances.  If associated
with the forward shock, this electron temperature corresponds to
an expansion speed of $\sim 1600 {\rm\ km\ s}^{-1}$ assuming
temperature equilibration between electrons and ions (and a higher
speed if equilibration has not yet been reached).

%%%%%%%%%%%%%%%%%%%%%%%%%% Figure 5 %%%%%%%%%%%%%%%%%%%%%%%%%%%%%%
\begin{figure}[t]
\centerline{\includegraphics[width=3.6in]{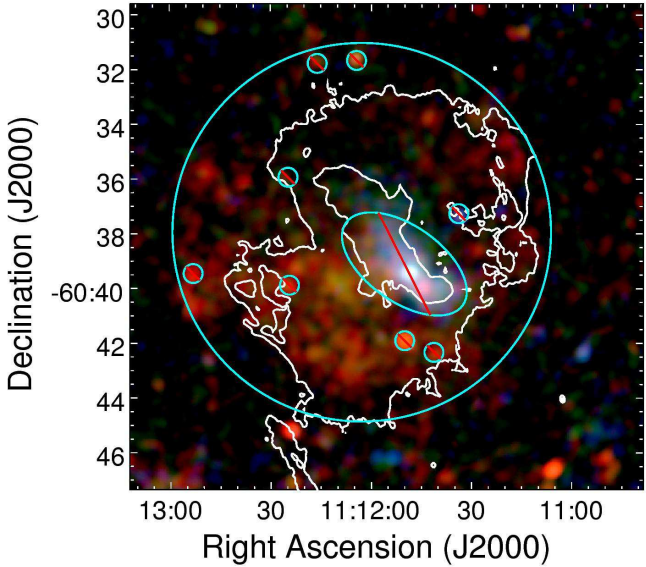}}
\caption{
\xmm\ MOS2 image of \msh, with radio contours from Figure 1. The image
has been exposure-corrected,
and red/green/blue corresponds to the energy band 0.6-1.5/1.5-2.5/2.5-5.0
keV. The large circle indicates the region from which the SNR spectrum
was extracted. Events from the inner ellipse were ignored to eliminate
the contribution from the PWN, and those from the small circles with
red slashes were
ignored to reduce point source contributions. The same inner ellipse
was used to extract the spectrum of the PWN.
}
\end{figure}
%%%%%%%%%%%%%%%%%%%%%%%%%%%%%%%%%%%%%%%%%%%%%%%%%%%%%%%%%%%%%%%%%%

Based on the volume emission measure for the lower-temperature
component, and assuming a thin-shell morphology for the SNR, with
a shell thickness of $R/12$ corresponding to a shock compression
ratio of 4, the estimated preshock density is $n_H = 0.04 \eta^{1/2}
d_5^{-1/2} f^{-1/2} {\rm\ cm}^{-3}$, where $\eta$ is the fraction
of the X-ray emission that is not associated with point sources in
the field, and $f$ is the fraction of the remnant shell actually
seen in X-rays. The observed variation in brightness around the SNR
shell suggests possible density variations of a factor of two or
so, and we consider the range $0.02 - 0.08 {\rm\ cm}^{-3}$ in
modeling the dynamical evolution in Section 5.  We note that the
higher temperature component provides very nearly the same density
estimate as that from the low temperature component.

The mass swept up by the SNR under such a scenario is then
$M_{sw} = 0.5 \eta^{1/2} f^{1/2} d_5^{5/2}M_\odot.$ Given that the
system morphology appears to show evidence that the PWN has
undergone an interaction with the SNR reverse shock, this low
value for the swept-up mass would appear to suggest a rather
low ejecta mass for the supernova. The inferred density, combined
with the ionization timescale from the spectral fits, indicates
an age of $5800 \eta^{-1/2} d_5^{1/2} f^{1/2}{\rm\ yr}$, which
should be considered an upper limit since portions of the SNR
shell that are not detected in X-rays presumably have lower 
density values.

%%%%%%%%%%%%%%%%%%%%%%%%%% Figure 6 %%%%%%%%%%%%%%%%%%%%%%%%%%%%%%
\begin{figure}[t]
\centerline{\includegraphics[width=3.4in]{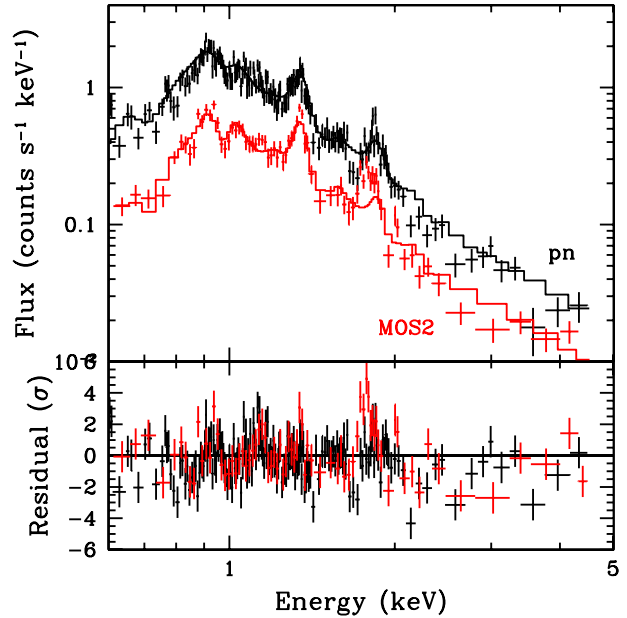}}
\caption{\xmm\ spectra from the SNR shell of \msh, with pn data shown
in black and MOS2 data shown in red.
The best-fit {\sc vpshock} model is shown as a histogram.}
\end{figure}
%%%%%%%%%%%%%%%%%%%%%%%%%%%%%%%%%%%%%%%%%%%%%%%%%%%%%%%%%%%%%%%%%%

\section{Gamma-ray Observations}

\subsection{\fermi\ LAT}

We investigated $\gamma $-ray events acquired from the region
surrounding \msh\ with the Fermi-LAT during the period 2008
August 5 to 2011 March 14. Standard event selection was applied,
using ``Source" class events with zenith angles less than 100 degrees
to minimize the portion of the Earth limb in the LAT field of view
(Abdo et al. 2009). The Pass 7 version 6 instrument response
functions were used, along with standard analysis tools available from the
Fermi Science Support Center (version v9r21p0).

Standard background models were used to account for both Galactic
and extragalactic diffuse emission as well as instrumental background.
Sources within 10\degr\ of \msh, with a statistical
significance larger than $5 \sigma$ above the background, are
extracted from the two-year Fermi-LAT Second Source Catalog (Abdo
et al.  2011), except for 2FGL J1112.1-6040 which corresponds to
the $\gamma$-ray emission of \msh. The flux parameters for all the
point-like sources less than 5\degr\ from \msh\ are left free in
the likelihood fit while the spectral parameters of other sources
are fixed at the 2FGL values.
The mapcube file
\emph{ring\_2year\_P76\_v0.fits} was used to describe the Galactic
$\gamma -$ray emission, and the isotropic component was modeled
using the \emph{isotrop\_2year\_P76\_source\_v0.txt} table.  Data
analysis details follow those in Castro \& Slane (2010) and Grondin
et al.  (2011).

Using the binned maximum likelihood analysis package {\sl gtlike},
the Test Statistic, $TS$, (for which the significance is approximately
$(TS)^{1/2} \sigma$; Mattox et al 1996, Equation 20) for \msh\ was
found to be 432 when the source was analyzed in the energy range
from 100 MeV to 100 GeV. This value of the Test Statistic corresponds
to a 21 sigma detection of the source. A map of the test statistic
as a function of position, used to localize the best fit position
of \msh, is shown in Figure 7.  Overlaid on this plot are blue
contours corresponding to ATCA observation of \msh\ and white
contours corresponding to the LAT significance levels of $\sim$ 13,
15, and 17 $\sigma$.

%%%%%%%%%%%%%%%%%%%%%%%%%% Figure 7 %%%%%%%%%%%%%%%%%%%%%%%%%%%%%%
\begin{figure}[t]
\centerline{\includegraphics[width=3.4in]{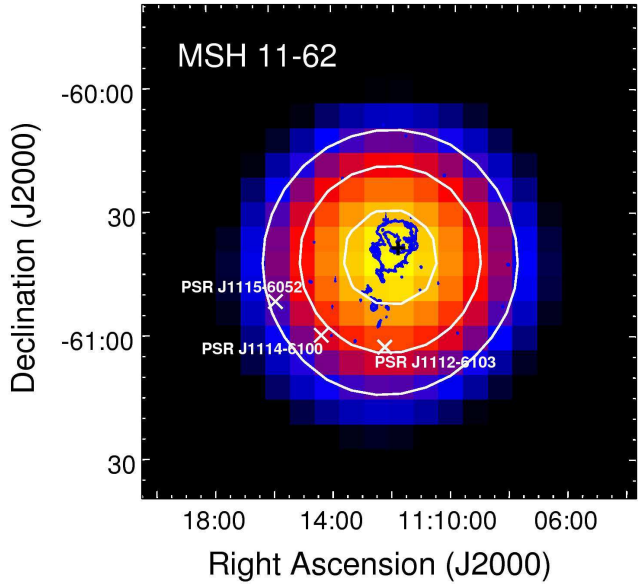}}
\caption{
$TS$ map of \fermi\ LAT emission from \msh. Blue contours
correspond to the radio emission, and white contours represent the
13, 15,  and 17 $\sigma$ significance levels of the flux. Positions of
nearby pulsars with $\dot{E} > 10^{32}{\rm\ erg\ s}^{-1}$ are indicated
(see discussion in text).
}
\end{figure}
%%%%%%%%%%%%%%%%%%%%%%%%%%%%%%%%%%%%%%%%%%%%%%%%%%%%%%%%%%%%%%%%%%

We also indicate the positions of the three
known pulsars within a $45^\prime$ radius from the centroid of the
LAT source. If each of these pulsars converts its entire spin-down
power into $\gamma$-rays in the LAT band, they could account for
no more than 6\% of the observed flux, and we thus ignore their
contributions in the subsequent analysis.  The nearest pulsar with
a high $\dot{E}/d^2$ ratio is PSR~J1105$-$6107, located $55^\prime$
from the centroid of the $\gamma$-ray emission, well beyond the
border of Figure 7; we consider an association with this source to
be very unlikely.

The source spectrum was extracted using both front and back events,
over the energy range 0.1-100 GeV. The spectrum is shown in Figure
8, and is well-described by a power law with $\Gamma = 1.2 \pm 0.2$
accompanied by an exponential cutoff with $E_{cut} = 3.2 \pm 0.9$~GeV
that improves the fit by $\sim 6.8 \sigma$.
The associated flux is $F(\geq 100 {\rm\ MeV}) = (6.3 \pm 1.2 \pm 3.2)
\times 10^{-8}{\rm\ photons\ cm}^{-2}{\rm\ s}^{-1}$ [or 
$(8.0 \pm 0.6 \pm 4.2) \times 10^{-11}{\rm\ erg\ cm}^{-2}{\rm\ s}^{-1}$],
where the quoted uncertainties are statistical and systematic,
respectively.

Two main systematic uncertainties can affect the LAT flux estimation
for a point source in the Galactic plane: uncertainties on the
Galactic diffuse background and on the effective area. The dominant
uncertainty at low energy comes from the Galactic diffuse emission,
which we estimated by changing the normalization of the Galactic
diffuse model artificially by 6\% as done in Abdo et al. (2010).
The second systematic is estimated by using modified IRFs whose
effective areas bracket those of our nominal Instrument Response
Function (IRF). These bracketing IRFs are defined by envelopes above
and below the nominal energy dependence of the effective area by
linearly connecting differences of (10\%, 5\%, 20\%) at log(E) of
(2, 2.75, 4), respectively.

%%%%%%%%%%%%%%%%%%%%%%%%%% Figure 8 %%%%%%%%%%%%%%%%%%%%%%%%%%%%%%
\begin{figure}[t]
\centerline{\includegraphics[width=3.4in]{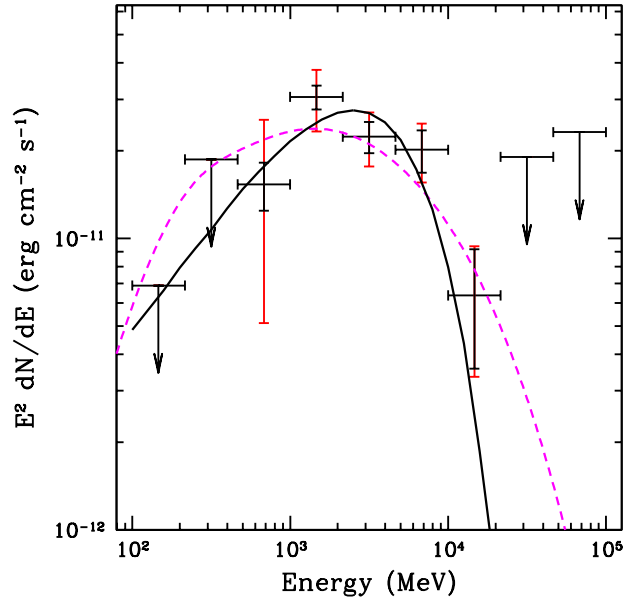}}
\caption{
\fermi\ LAT spectrum of \msh. Black error bars correspond to
statistical errors, while red error bars include systematic errors
as well. The solid curve corresponds to the best-fit model for
a power law with an exponential cutoff. The dashed magenta curve
corresponds to a pion-decay model assuming a power law proton
spectrum with an exponential cutoff. See Section 5.2 for discussion.
}
\end{figure}
%%%%%%%%%%%%%%%%%%%%%%%%%%%%%%%%%%%%%%%%%%%%%%%%%%%%%%%%%%%%%%%%%%

The source 2FGL J1112.1-6040, from the LAT 2$^{nd}$ year catalog,
was identified with a possible association with \msh (Abdo et al.
2011).  The 2FGL flux agrees with our measurements as well, but
both are significantly lower than that of 3EG J1102-6103, whose
quoted spectral index is also significantly steeper than that for
\msh.  These differences are presumably associated with the
previously-unrecognized spectral cutoff.

We have searched for possible extension of the source using the
routine {\sc pointlike} (Kerr 2011), comparing the
test statistic $TS_{ext}$ for the extended hypothesis to that for
a point source hypothesis, $TS_p$, defining $TS_{ext}=TS-TS_p$. We
selected events converting in the front section of the tracker,
where the Point Spread Function (PSF) is the narrowest.  We fitted
the position of the source assuming iteratively a point source and
a Gaussian shape for which we fitted the extension.
We find that the best-fit centroid is located at 11h12m12s, $-$60\degr
40'12", in good agreement with the position of \msh, and that there
is no evidence for extent with significance greater than $3 \sigma$.
Computing a
significance map, the {\sc pointlike} routine determines the
likelihood gradient with spatial coordinates; this gradient has a
$1 \sigma$ width of 1.2 arcmin in each direction for \msh. The
significance contours thus provide an indirect way of characterizing
the statistical uncertainties on the position of the detected source.
A deeper observation is required to firmly address, with higher
confidence, whether or not the \fermi\ LAT source is extended.

\subsection{H.E.S.S. Upper Limits}

The position of \msh\ was covered in a survey of the inner Galactic
plane with the High Energy Spectroscopic System (H.E.S.S.) in
2004/2005. The approximate exposure time on the source was 9.1 hr
(Hoppe 2008), resulting in no reported detection of the
source. Based on this non-detection, we calculate a $2 \sigma$ upper
limit on the VHE flux at 1 TeV for \msh\ by scaling the H.E.S.S.
sensitivity for a $5 \sigma$ point source detection (1\% of the Crab
in 25 h under the assumption of a photon index of 2.6) to the actual
H.E.S.S.  exposure for the source.  We find $E^2 (dN/dE)_{E = 1
TeV} < 3.8 \times 10^{-12} {\rm\ erg\ cm}^{-2} {\rm\ s}^{-1}$,
where $N$ is the number of photons per unit area per second.

%%%%%%%%%%%%%%%%%%%%%%%%%% Figure 9 %%%%%%%%%%%%%%%%%%%%%%%%%%%%%%
\begin{figure}[t]
\centerline{\includegraphics[width=3.4in]{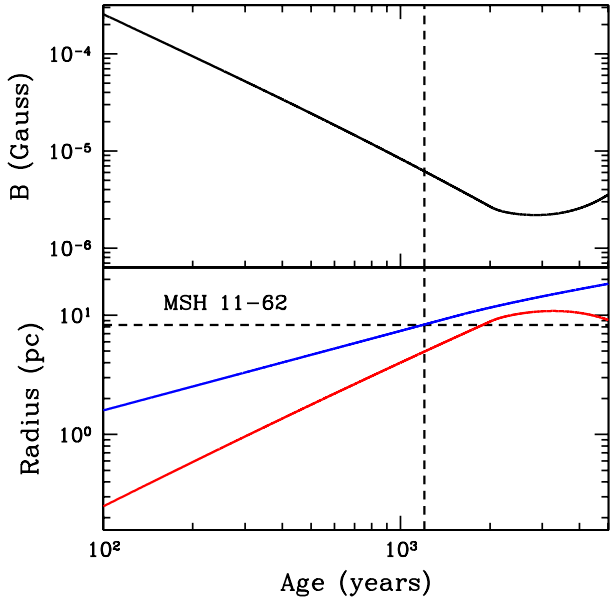}}
\caption{
Upper: Time evolution of the PWN magnetic field. Lower: Evolution of
the SNR (blue) and PWN (red) radii as a function of time. The
observed radius of \msh\ is indicated, and the vertical line
indicates the age at which this radius is reached. See Section 5.1 for
model details.
}
\end{figure}
%%%%%%%%%%%%%%%%%%%%%%%%%%%%%%%%%%%%%%%%%%%%%%%%%%%%%%%%%%%%%%%%%%

\section{Analysis and Discussion}

\subsection{Evolutionary State}

The large-scale evolution of a composite SNR depends on the mechanical
energy of the explosion, the density of the ambient medium, the mass
of the supernova ejecta, and the spin-down power of the pulsar. Additional
parameters of considerable importance are the braking index and spin-down
timescale of the pulsar, along with the fraction of the spin-down power
that appears as magnetic flux.

Figure 9 shows an illustrative model for the evolution of \msh\
using parameters for a scenario described in Section 5.3.  For these
parameters, the forward shock (FS) of the SNR (blue curve) reaches
the observed radius of \msh\ 
($R_{SNR} \approx 8.3 d_5 {\rm\ pc}$)
at an age of $\sim 1.2$~kyr. By this
time, the reverse shock has begun to approach the PWN
(for which the model yields $R_{PWN} \approx 5 {\rm\ pc}$,
or an angular diameter of $\approx 6.9$~arcmin at a distance
of 5 kpc),
as indicated
by the subsequent compression of the PWN radius (red curve).  This
is consistent with the somewhat distorted morphology of \msh, which
appears to indicate that the RS has propagated most rapidly along
the NW/SE direction.  In addition, Figure 9 shows that the magnetic
field in the PWN has declined to a value of $\sim 6 \mu{\rm G}$.

The broadband emission observed from \msh\ places considerable
constraints on the underlying properties of the composite system.
Of particular importance is our understanding of the observed
$\gamma$-ray emission, which could originate from the pulsar, the
PWN, or the SNR itself. Here we consider models for each scenario
using constraints derived from the radio and X-ray
observations of \msh.
In these model investigations, we have varied input parameters to
investigate the broad effects they have on the broadband spectrum,
and have then fine-tuned parameters to provide reasonable agreement
with the data. The resulting models are not statistical fits to the
observed emission, and were not optimized by a likelihood analysis
over the large parameter space.

\subsection{Gamma-rays from the SNR}

Particle acceleration at the forward shocks of SNRs is known to
produce extremely energetic particles that can produce $\gamma$-rays
through inverse-Compton (IC) scattering of ambient photons by
energetic electrons, and/or through the decay of neutral pions
produced in energetic proton-proton collisions. Thus, one interpretation
for the $\gamma$-ray emission from \msh\ is that it originates from
the SNR itself, and is a signature of particle acceleration by the
remnant. We consider two cases, one in which the $\gamma$-ray
emission is dominated by hadrons ($\pi^0$-decay), and one in which
it is dominated by leptons (IC scattering).

For the hadronic-dominated scenario, we assume that the mechanical
energy of the SNR is $E_{SNR} = 10^{51}{\rm\ erg}$ and that 50\% of this
energy is converted to energetic particles whose spectrum is described
by a power law with an exponential cutoff. We assume an electron-to-proton
ratio $k_{e-p} = 10^{-2}$, and that the spectral index is the same
for protons and electrons. Using $d = 5$~kpc, we adjust the spectral
index, density, and cutoff energy for the proton spectrum to match
the Fermi LAT spectrum, while also adjusting the magnetic field
strength to reproduce the SNR radio emission with synchrotron emission
from the electron spectrum. We find a spectral index of 1.8 and an
exponential cutoff energy of 70~GeV for the protons (Figure 10,
top), with an ambient density $n_0 = 6.8 {\rm\ cm}^{-3}$.  The
associated $\gamma$-ray emission from IC scattering of the cosmic
microwave background (CMB) is negligible.  The postshock magnetic
field required by the radio emission is $\sim 10 \mu$G. Here we
have used the measured 1.4~GHz flux from the shell, and assumed a
spectral index range of $\alpha_r = 0.4 - 0.7$ typical of radio
emission from shell-type SNRs.

%%%%%%%%%%%%%%%%%%%%%%%%%% Figure 10 %%%%%%%%%%%%%%%%%%%%%%%%%%%%%%
\begin{figure}[t]
\centerline{\includegraphics[width=3.2in]{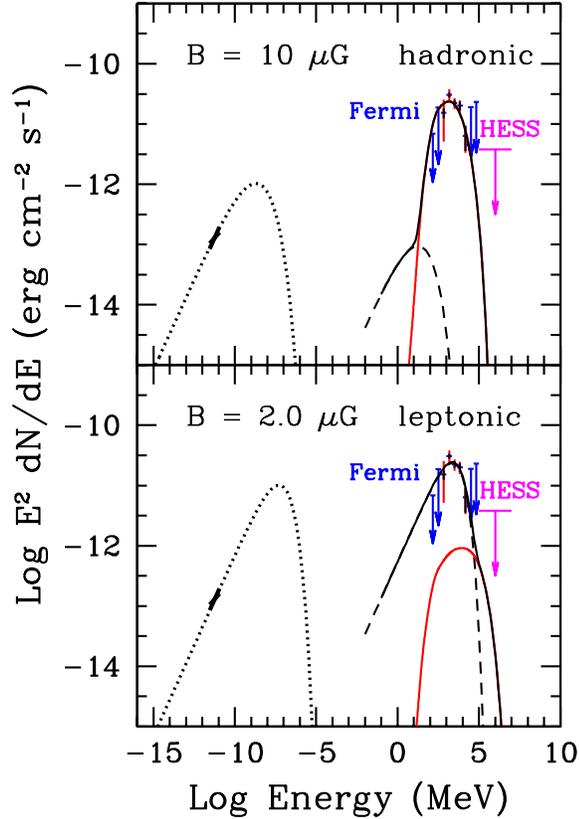}}
\caption{
Models for the broadband emission from \msh\ under the assumption
of dominant hadronic (upper) and leptonic (lower) processes for
the observed $\gamma$-ray emission. The dotted curves correspond
to synchrotron emission, and the dashed curves correspond to IC
emission. The red curve represents $\gamma$-ray emission from pion
decay, and the solid black line represents the summed $\gamma$-ray
components.
}
\end{figure}
%%%%%%%%%%%%%%%%%%%%%%%%%%%%%%%%%%%%%%%%%%%%%%%%%%%%%%%%%%%%%%%%%%

The required proton cutoff energy under this scenario is rather low
given that we expect energies well in excess of $10^{12}$~eV to be
produced in young SNRs, but it is possible that a low shock speed
has resulted in a low maximum energy (e.g., Reynolds 2008).  Of
potentially greater concern is that the preshock density required
under this scenario is more than two orders of magnitude larger
than that inferred from the X-ray spectrum. Similar discrepancies
have been identified for several $\gamma$-ray emitting SNRs associated
with OH masers (which imply an interaction between the SNRs and
surrounding molecular clouds, thus making $\pi^0$-decay a likely
production mechanism), and it has been suggested that this could
be the result of a clumpy SNR shell in which only the lower-density
interclump medium is responsible for the X-ray emission while all
of the gas participates in the $\gamma$-ray production (Castro \&
Slane 2010).  No maser emission has been reported for \msh, and the
remnant is not known to be interacting with dense material. Thus,
while this scenario cannot be ruled out, particularly since the
production of masers requires a rather narrow range of physical
conditions, it does not have strong supporting evidence outside of
the $\gamma$-ray emission.  Searches for evidence of interactions
between \msh\ and molecular clouds are of particular importance to
further address this picture.

For at least some SNRs, the observed $\gamma$-ray emission appears
to be dominated by IC emission from an energetic population of
electrons (e.g., Ellison et al. 2010). We have considered such a
scenario for \msh\ by fixing the ambient density at the value implied
by the X-ray measurements, and allowing the normalization and shape
of the electron spectrum to vary.  We find that for a distance of
5~kpc, the total electron energy required to produce the $\gamma$-ray
emission through IC scattering of CMB photons is $8 \times 10^{49}{\rm\
erg}$. The magnetic field required to fit the radio emission from
the SNR shell is $B = 2.0 \mu$G.  The model, shown in Figure 10
(bottom), requires a super-exponential cutoff of the electron
spectrum
\begin{equation}
N_e \propto E_e^{-\alpha_e} \exp\left[-\left(\frac{E_e}
{E_{cut,e}}\right)^b\right]
\end{equation}
Model parameters are given in Table 2.  In particular, we find
$E_{cut,e} = 0.9$~TeV, which is quite low given the rather low
magnetic field.  Since all realistic acceleration models that produce
energetic nonthermal electrons in SNRs also accelerate ions, the
total energy budget implied by these results would exceed the
available energy of the SNR blast wave for $k_{e-p} \sim 10^{-2}$
if the remnant distance is 5~kpc.

%%%%%%%%%%%%%%%%%%%%%% Table 1 %%%%%%%%%%%%%%%%%%%%%%%%%%%%%%%%

\begin{deluxetable}{lcc}
\tablecolumns{3}
\tabletypesize{\scriptsize}
\tablewidth{0pc}
\tablecaption{Model Parameters: $\gamma$-rays from SNR}
\tablehead{
\colhead{Parameter} &
\colhead{Hadronic Model} &
\colhead{Leptonic Model} 
}
\startdata
$d$ (fixed)		& 5 kpc 			& 5 kpc \\
$n_0$			& $6.8 {\rm\ cm}^{-3}$ 	& $0.04 {\rm\ cm}^{-3}$ \\
$\alpha_e$ ($ = \alpha_p$)		& 1.8 		& 1.8 \\
$E_{cut,e}$ 		& 50 GeV			& 0.9 TeV \\
$E_{cut,p}$		& 70 GeV		& 1 TeV \\
$b^{\rm a}$		& 1.0		& 2.0 \\
$E_e^{\rm b}$ 		& $5 \times 10^{48}{\rm\ erg}$	& $8 \times 10^{49}{\rm\ erg}$ \\
$E_p^{\rm c}$ 		& $5 \times 10^{50}{\rm\ erg}$	& $8 \times 10^{51}{\rm\ erg}$ \\
$k_{e-p}$ (fixed)	& $10^{-2}$	& $10^{-2}$  \\
$B$		& $10.0 \mu$G 			& $2.0  \mu$G\\
\enddata
\tablecomments{
\\
a) Curvature index on exponential cut-off\\
b) Total electron energy \\
c) Total proton energy
}
\end{deluxetable}

%%%%%%%%%%%%%%%%%%%%%%%%%%%%%%%%%%%%%%%%%%%%%%%%%%%%%%%%%%%%%%%%%%%%%%%%

For the smallest distance allowed by HI absorption measurements,
an electron-to-proton ratio $k_{e-p} > 0.7$ is required for the
entire particle energy budget to not exceed $5 \times 10^{51}{\rm\
erg}$.  Such high required values of $k_{e-p}$, along with the low
cutoff energy and extremely sharp spectral cutoff appear problematic.
Thus, we conclude that a leptonic scenario in which the observed
$\gamma$-ray emission from \msh\ is dominated by IC emission is
ruled out for all but the smallest possible distances, and still
problematic at such small distance values.

\subsection{Gamma-rays from the PWN}

The X-ray and radio observations of \msh\ clearly identify a powerful
PWN in the system. The electrons responsible for the X-ray and
radio emission also produce $\gamma$-ray emission through inverse
Compton scattering. To investigate the scenario in which the Fermi LAT
emission is produced entirely by the PWN, we calculate the evolution
of the composite system using the model of Gelfand, Slane, and
Zhang (2009). We assume expansion of the SNR into a uniform medium
of density $n_0$, driven by an ejecta mass $M_{ej}$. The PWN expansion
is powered by the input of a central pulsar with a spin-down
power that evolves as
\begin{equation}
\dot{E}(t) = \dot{E}_0 \left[1 + \frac{t}{\tau_0}\right]^{-\frac{p+1}{p-1}}
\end{equation}
where $\tau_0$ is the initial spin-down timescale for the pulsar,
and $p$ is the braking index.  The PWN expansion is confined by the
surrounding ejecta. Particles are injected into the nebula with a
power law spectrum, and evolved through radiative and adiabatic
losses. The emission spectrum is then calculated based upon the
evolved magnetic field strength of the nebula at the age required
to reproduce the observed radius of the SNR (Figure 9). The ambient
radiation field was assumed to be a combination of the cosmic
microwave background (CMB) and infrared photons from local dust.

%%%%%%%%%%%%%%%%%%%%%%%%%%%%%% Figure 11 %%%%%%%%%%%%%%%%%%%%%%%%%%%%%%
\begin{figure}[t]
\centerline{\includegraphics[width=3.2in]{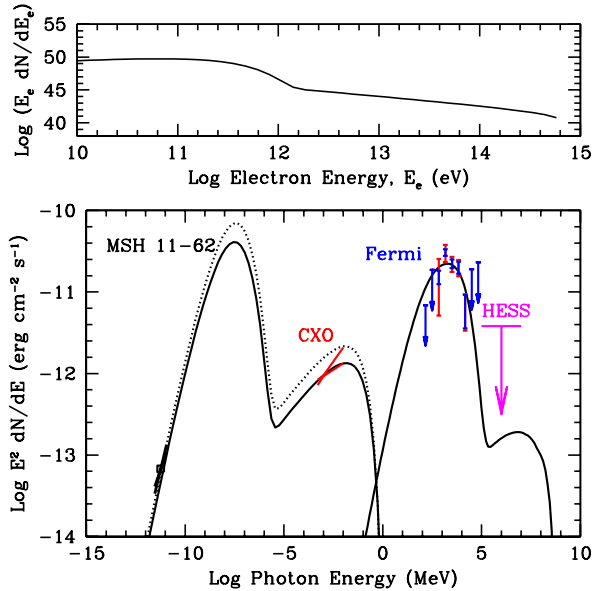}}
\caption{
Models for the nonthermal emission from \msh\ (lower panel) under
the assumption that the $\gamma$-ray emission is produced entirely
by the PWN.
Dashed, solid, and dotted lines correspond to $n_0 = 0.02$, $0.04$, and
$0.08 {\rm\ cm}^{-3}$.
The upper panel shows the associated electron distribution, which
is modeled as an evolved Maxwellian distribution with power
law tail.
}
\end{figure}
%%%%%%%%%%%%%%%%%%%%%%%%%%%%%%%%%%%%%%%%%%%%%%%%%%%%%%%%%%%%%%%%%%

The parameters of the model are summarized in Table 3. 
We fixed the distance at 5~kpc and assumed $E_{SNR} = 10^{51}{\rm\
erg}$.  We sampled values for $n_0$ in the range $0.02 - 
0.08 {\rm\
cm}^{-3}$ based on the estimates provided from the X-ray emission
(see Section 3.4), and determined that the broadband spectrum is
not highly sensitive to the ambient density, with flux values varying
by only $\sim 30$\% in the X-ray band (see Figure 11).
Braking index values over the range $p = 2 - 3$, and
ejecta mass values between $1.5 - 8.5 M_\odot$, produce flux
variations similar in magnitude to those observed for the above
range in $n_0$, and we thus fixed these quantities at the values
shown in Table 3. The observed radio flux provides an upper limit
for the minimum electron energy; lower values of $E_{e,min}$ do not
appreciably change the spectrum.  Similarly, the X-ray spectrum
provides a lower limit for the maximum electron energy. Higher
values of $E_{e,max}$ allow the photon spectrum to extend to higher
energies, but the current data provide no additional constraints
on this.

%%%%%%%%%%%%%%%%%%%%%% Table 3 %%%%%%%%%%%%%%%%%%%%%%%%%%%%%%%%

\begin{deluxetable}{lcc}
\tablecolumns{3}
\tabletypesize{\scriptsize}
\tablewidth{0pc}
\tablecaption{Model Parameters: $\gamma$-rays from PWN or Pulsar}
\tablehead{
\colhead{Parameter} &
\colhead{$\gamma$-rays from PWN} &
\colhead{$\gamma$-rays from Pulsar}
}
\startdata
{\it Input:} && \\
$d$ (fixed)			& 5 kpc 			& 5 kpc \\
$E_{SNR}$ (fixed)	& $10^{51}$ erg 	& $10^{51}$ erg \\
$M_{ej}$ (fixed)	& $2.5 M_\odot$ 		& $2.5 M_\odot$ \\
$n_0$ (fixed)		& $0.04 {\rm\ cm}^{-3}$ 	& $0.04 {\rm\ cm}^{-3}$ \\
$p$ (fixed)		& 3.0 			& 3.0  \\
$\tau_0$		& 1000 yr 		& 1000 yr (fixed)\\
$\dot{E_0}$		& $4 \times 10^{39} {\rm\ erg\ s}^{-1}$ & $3 \times 10^{38} {\rm\ erg\ s}^{-1}$\\
$\eta_e^{\rm a}$	& 0.999 		& 0.995\\
$\eta_B^{\rm b}$	& 0.001 		& 0.005 \\
$E_{e,min}$		& 0.1 GeV 		& 1 MeV \\
$E_{e,max}$		& 900 TeV 		& 8.5 PeV\\
$E_p^{\rm c}$		& 100 GeV & \\
$f_{PL}^{\rm d}$	& .1\% & \\
$E_b^{\rm e}$			& 			&4.4 GeV \\
$\alpha_e$		& 2.33  & 		0.0 \\
$\alpha_{e,2}$		& 			& 2.7 \\
$T_{IR}^{\rm f}$  	& 25K 			& 25K (fixed)\\
$(W_{IR}/W_{CMB})^{\rm g}$	& $2 W_{CMB}$ 		& $2 W_{CMB}$ (fixed)\\
&& \\
{\it Derived:} && \\
Age			& 1200 yr 			& 1300 yr\\
$B_{PWN}$		& $6.1 \mu$G 			& $7.3 \mu$G\\
\enddata
\tablecomments{
\\
a) Fraction of energy in electrons\\
b) Fraction of energy in magnetic field\\
c) Peak energy of Maxwellian electron component \\
d) Fraction of energy in power law component \\
e) Power law break energy \\
f) Effective temperature of IR photon field \\
g) Ratio of IR energy density relative to CMB\\
}
\end{deluxetable}

%%%%%%%%%%%%%%%%%%%%%%%%%%%%%%%%%%%%%%%%%%%%%%%%%%%%%%%%%%%%%%%%%%%%%%%%

With the above parameters fixed, we varied values for the pulsar
spin-down timescale, $\tau_0$, the initial spin-down energy,
$\dot{E}_0$, the electron spectral index, $\alpha_e$, and the
fraction of the pulsar wind energy in the form of magnetic fields
and particles, $\eta_B$ and $\eta_e$ respectively.  We found that
it was not possible to reproduce the significant $\gamma$-ray
emission in the Fermi LAT band with simple power law or broken power
law models for the particle spectrum under these assumptions. We
then investigated a model consisting of a Maxwellian electron
distribution with a power law tail, corresponding to spectra observed
in PIC simulations of relativistic shocks (Spitkovsky 2008),
considering a range of values for the energy of the Maxwellian peak
and for the fraction of the total electron energy found in the power
law component.  The temperature and energy density of the IR photon
field were adjusted slightly to tune the agreement with the data.
The final values listed along with the other model parameters 
in Table 3 are within reasonable expectations
for this region of the Galaxy (see Strong et al. 2000).

As shown in Figure 11, we find that the broadband
emission from \msh\ can be reproduced to reasonable fidelity for
the parameters shown in Table 3. The upper panel of the Figure
contains the electron spectrum, and the lower panel presents the
synchrotron and IC emission along with the observed data. The radii
for the SNR and PWN are shown in Figure 9, along with the PWN
magnetic field. The age required to match the observed 
SNR
radius is
$\sim 1.2$~kyr. The magnetic field of the PWN has declined to a
value of $\sim 6 \mu$G, and the PWN radius is approaching the stage
at which it encounters the SNR reverse shock, as indicated by the
subsequent decline in $R_{PWN}$. 
This appears qualitatively consistent with the X-ray and radio
morphology, as discussed in Section 5.1, however the narrowest
portion of the PWN is considerably smaller than indicated in Figure
9.  The very distorted shape of the nebula suggests that the reverse
shock has propagated asymmetrically in \msh, which is not a scenario
that can be treated in our spherical model.  

The magnetic field in this scenario is far below the minimum-energy
value calculated from radio measurements, indicating that the nebula
is particle-dominated.
We note that the $\sim 65 \mu$G field estimated by Harrus, Hughes,
\& Slane (1998) was based on the assumption that the minimum-energy
conditions hold in \msh. Additional estimates based on the interpretation
of the spectral break between the radio and X-ray bands as being an
evolutionary effect, and on the pressure in the PWN being in equilibrium
with that of the thermal gas in the SNR, led to even higher estimated
values of the magnetic field. However, no modeling was performed to 
demonstrate which, if any, of those assumed conditions provide a 
self-consistent picture of the composite system. Here we have carried
out such modeling efforts and find that a particle-dominated nebula
is required, with the magnetic field at the current epoch being well 
below the equipartition value.
Similar results are obtained for the evolved
nebulae in W44 (Bucciantini et al. 2011), G0.9+0.1, and G338.3-0.0
(Fang \& Zhang 2010). We also note that even though different values for
the assumed distance can lead to different values of the model parameters
(see below), the observed X-ray spectral index constrains the magnetic field
to be low, and the overall magnetization to be low as well.

According to this model, $\gamma$-ray emission in the Fermi LAT band
is produced entirely by electrons in the Maxwellian peak of the
spectrum. This is consistent with results for Vela X (de Jager,
Slane, \& LaMassa 2008), HESS J1640$-$465 (Slane et al. 2010), and
several other PWNe (Fang \& Zhang 2010) that suggest a low-energy
electron component distinct from the power law component 
(although for Vela X this was modeled as an additional power 
law component). 
For \msh, we find a mean Lorentz factor of $\gamma = 2 \times 10^5$
for the electrons in the Maxwellian component, with only 0.1\% of
the total energy in the power law tail.

We note that the initial spin-down power required for this model,
$\dot{E}_0 = 4 \times 10^{39}{\rm\ erg\ s}^{-1}$, is uncomfortably
large. For the assumed braking index $p = 3$, this would lead to a
value of $\dot{E} = 8.3 \times 10^{38}{\rm\ erg\ s}^{-1}$ at the
current epoch -- higher than that for any known pulsar. This would
imply $L_x/\dot{E} \sim 10^{-5}$ for the pulsar, which is extremely
low.

Given that the distance to \msh\ is not well known, the numerical
values from the model above are quite uncertain. Changes in the
assumed distance can be accommodated, at least in part, by adjusting
the spin-down power, for example. However, larger distance values
lead to even higher values of $\dot{E}$.  Reasonable fits to the
broadband spectrum can also be obtained for a distance as low as
2~kpc, yielding a current spin-down power of $\sim 2 \times
10^{38}{\rm\ erg\ s}^{-1}$. 
While still extremely high, this
is similar to the value for PSR~J2022$+$3842, a pulsar recently
discovered in G76.9$+$1.0 (Arzoumanian et al. 2011).  This pulsar
is characterized by an extremely low conversion efficiency of
spin-down power into X-rays as well, but the value ($L_x/\dot{E}
\sim 6 \times 10^{-5}$) is still much higher than that inferred for
the pulsar in \msh\ if the Fermi LAT emission is being produced by
the PWN. 
We conclude that it is possible for the observed 
$\gamma$-ray emission from \msh\ to be produced by the PWN, but
that this would require a very significant amount of energy in
the Maxwellian peak of the electron spectrum, leading to spin-down
parameters for the pulsar that seem unlikely.

\subsection{Gamma-rays from the Pulsar}

\fermi\ observations have clearly established that pulsars comprise
a significant fraction of the Galactic population of GeV point
sources.  The associated spectra are characterized by a power law
with an exponential cutoff, with $E_{\rm cut} \sim 1 - 6$~GeV, and
$\Gamma \sim 1 - 2$ (Abdo et al. 2010). This is very similar to the
spectrum observed for \msh. Being associated with a composite SNR,
for which our X-ray observations identify the underlying neutron
star, it is quite possible that the observed $\gamma$-ray emission
originates with the pulsar itself.

The ratio of the X-ray flux from the point source in \msh\ to the
observed $\gamma$-ray flux is $F_x/F_\gamma \sim 5 \times 10^{-3}$.
This value is typical of that for pulsars detected by the \fermi\
LAT, although there is a large spread in this ratio for the population.

%%%%%%%%%%%%%%%%%%%%%%%%%% Figure 12 %%%%%%%%%%%%%%%%%%%%%%%%%%%%%%
\begin{figure}[t]
\centerline{\includegraphics[width=3.2in]{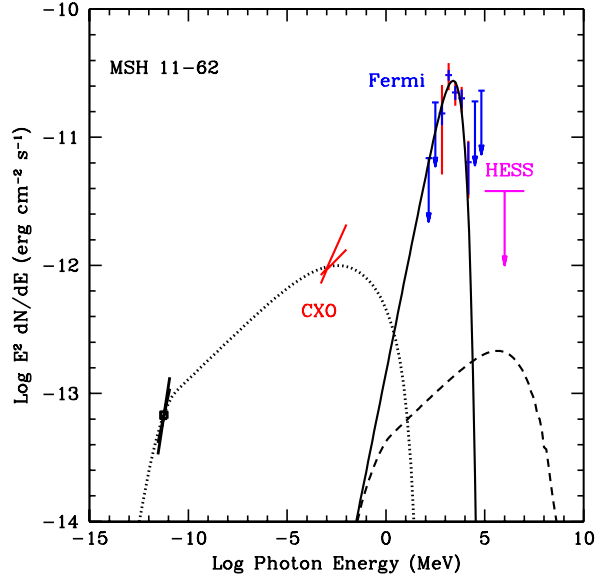}}
\caption{
Model for the PWN emission from \msh\ under the assumption
that the $\gamma$-ray emission arises from the pulsar. Parameters
for the model are given in Table
3.
}
\end{figure}
%%%%%%%%%%%%%%%%%%%%%%%%%%%%%%%%%%%%%%%%%%%%%%%%%%%%%%%%%%%%%%%%%%

Using the approximate relationship $L_x \approx 10^{-3} \dot{E}$
derived from a variety of studies of X-ray PWNe and their associated
pulsars, the spectrum of the PWN in \msh\ suggests an underlying
pulsar with $\dot{E} \sim 10^{37} d_5^2 {\rm\ erg\ s}^{-1}$.  We
note, however, that there are large variations in the observed
relation between $L_x$ and $\dot{E}$ (e.g., Vink, Bamba, \& Yamazaki
2011). More than 85\% of the pulsars detected by the \fermi\ LAT,
and all of those pulsars discovered in the $\gamma$-ray band, have
lower values of $\dot{E}$.  However, of currently reported pulsars
discovered in $\gamma$-rays, only one has a \fermi\ LAT flux lower
than that associated with \msh. Thus, the lack of identified
pulsations in any band is not strong evidence that the emission
from \msh\ is not associated with the pulsar itself. The marginal
evidence for extension of the $\gamma$-ray emission, which would
rule out the pulsar as the source of the emission, is of particular
importance in this context, and requires revisiting in the future
when additional data from this source have been obtained.

If we assume that the \fermi\ LAT emission arises entirely from the
pulsar in \msh, the PWN radio and X-ray emission can then be
reproduced by a broken power law electron spectrum with $\dot{E}_0
\sim 3 \times 10^{38}{\rm\ erg\ s}^{-1}$ and an age of $\sim 1300$~yr,
at which time the ambient density $n_0 = 0.04{\rm\ cm}^{-3}$ inferred
from the SNR X-ray emission yields the observed radius for the SNR
assuming a distance of 5 kpc.  Here, as in Section 5.3, we have
fixed the values for $d$, $M_{ej}$, $n_0$, and $p$. In addition,
we have used the same model values for the IC photon fields as were
used in Section 5.3, although these are largely irrelevant here
since the IC emission falls well below the observed $\gamma$-ray
emission. We have varied the remaining parameters in Table 3 in
order to provide an approximate match to the observed radio and
X-ray spectra.  The resulting model spectrum is shown in Figure 12,
and the parameter values for the model are listed in Table 3.  The
PWN magnetic field strength is $\sim 7 \mu$G, a factor of 5 lower
than the minimum-energy field derived from radio observations, and
the spin-down power at the current epoch is $\dot{E} = 5.7 \times
10^{37}{\rm\ erg\ s}^{-1}$. This leads to $L_x/\dot{E} \approx 2
\times 10^{-4}$, somewhat low, but well within the observed range
for known pulsars.  In this model, the SNR RS has not yet begun to
impact the PWN, leaving the observed X-ray and radio morphology
unexplained.

We note that this model again leads to a particle-dominated nebula,
with the magnetic energy comprising only $\sim 4\%$ of the total
energy. Models that produce larger magnetic fields at the derived
age can be obtained by increasing $\eta_B$. However this leads to
a synchrotron-loss cutoff below the X-ray band.
Conversely, a better fit to the X-ray slope can be obtained by
further reducing the fraction of wind energy that appears as
magnetic flux. This, however, requires a much higher value of
$\dot{E_0}$ and results in an even lower value for the current
magnetic field strength in the PWN.
In any event, the low magnetization of the nebula appears
robust in these models, although we caution that 
our spherically-symmetric semi-analytical calculations do not
address more complex geometrical and magnetohydrodynamical effects
that surely complicate the magnetic field structure in PWNe.

\section{Conclusions}

Radio observations of \msh\ indicate that this is a composite SNR
containing a central PWN. X-ray observations confirm this, with
clear identifications of the PWN synchrotron emission as well as
the presence of an X-ray point source that most likely corresponds
to the pulsar driving the system. The X-ray spectral index in the
PWN increases with distance from the compact source, as expected
from synchrotron losses, and the strongly asymmetric morphology of the PWN
suggests an interaction with the SNR RS.

Observations with the \fermi\ LAT reveal distinct $\gamma$-ray
emission from \msh, but the limited angular resolution makes it
impossible to ascertain the origin of the emission directly. We
have considered three cases, with the $\gamma$-ray emission arising
from the SNR, the PWN, and the pulsar, respectively. Models for
each scenario can reproduce basic features of the observed broadband
nonthermal emission, but not without difficulties. If the emission
arises from the SNR, then a hadronic scenario requires an ambient
density that is a factor of $\sim 100$ higher than that implied by
the thermal X-ray emission, while a leptonic scenario requires that
an unreasonably large fraction of the SNR mechanical energy is
converted to relativistic particles unless the SNR distance is quite
low. Both models require very low energy cutoffs in the particle
spectra. If the emission arises from the PWN, then the input spectrum
from the pulsar requires a dominant Maxwellian component, with only
0.1\% of the energy in the accompanying power law tail, and an
extremely high pulsar spin-down power that appears inconsistent
with the observed X-ray and $\gamma$-ray flux values.

The most likely scenario is that the $\gamma$-ray emission arises from
the pulsar itself. While models for the PWN emission still imply a
somewhat high spin-down power, the lack of detected $\gamma$-ray
pulsations is not surprising given the relatively low flux of the
source. The weak evidence for extension in the $\gamma$-ray emission
is not sufficient to reject this scenario, but is of considerable
interest for additional study with future observations.

\acknowledgments

The authors thank the referee, Rino Bandiera, for his careful
review and helpful commentary on this manuscript.

NASA supported this work via grant numbers NRA 00-OSS-07/03500279
and NGT5-159. PS acknowledges support from NASA Contract NAS8-03060.
The ATCA is funded by the Commonwealth of Australia for operation
as a National Facility managed by CSIRO.

RR, ML-G, and SF have participated as members of the Fermi LAT
Collaboration, which acknowledges generous ongoing support from a
number of agencies and institutes that have supported both the
development and the operation of the LAT as well as scientific data
analysis.  These include the National Aeronautics and Space
Administration and the Department of Energy in the United States,
the Commissariat \`a l'Energie Atomique and the Centre National de
la Recherche Scientifique / Institut National de Physique Nucl\'eaire
et de Physique des Particules in France, the Agenzia Spaziale
Italiana and the Istituto Nazionale di Fisica Nucleare in Italy,
the Ministry of Education, Culture, Sports, Science and Technology
(MEXT), High Energy Accelerator Research Organization (KEK) and
Japan Aerospace Exploration Agency (JAXA) in Japan, and the
K.~A.~Wallenberg Foundation, the Swedish Research Council and the
Swedish National Space Board in Sweden.  Additional support for
science analysis during the operations phase is gratefully acknowledged
from the Istituto Nazionale di Astrofisica in Italy and the Centre
National d'\'Etudes Spatiales in France.

%-------------------------------------------------------------------------------
%-------------------------------------------------------------------

%\bibliographystyle{apj}                       %% AASTeX

%\bibliography{mn-jour,example1}     %% includes the journal abbrevs
%\bibliography{apj-jour,example1}    %% includes the journal abbrevs
%\bibliography{refs-msh}
\bibliographystyle{plain}

%%% REFERENCES

\end{document}